# What is the Mechanism Underlying 3-D Heisenberg-like Ferromagnetism Across the Compositional Metal-Insulator Transition in La$_{1-x}$Ca$_x$MnO$_3$ (0.18 ≤ x$_c$ ≤ 0.22)?

Wanjun Jiang[1*], X. Z. Zhou[1], K. Glazyrin[2], Y. Mukovskii[2], Gwyn Williams[1]
1: Department of Physics and Astronomy, University of Manitoba, Winnipeg, Manitoba, R3T 2N2 CANADA
2: Moscow State Steel and Alloys Institute, Moscow 119049 RUSSIA

Detailed measurements of the magnetic and transport properties of the two La$_{1-x}$Ca$_x$MnO$_3$ (x = 0.18, x = 0.20) single crystals straddling the compositional metal-insulator transition boundary (0.18 ≤ x$_c$ ≤ 0.22) are summarized. The analysis of magnetization/susceptibility data reveals the occurrence of a second order/continuous ferromagnetic-paramagnetic phase transition described not only by nearest neighbour, 3-D Heisenberg model exponents (γ = 1.387, β = 0.365, δ = 4.783), but also with comparable values of the critical amplitudes in both the insulating and the metallic samples. These results support the assertion that double exchange cannot be the underlying mechanism supporting ferromagnetism in this composition range, and arguments are presented that the relevant interaction is ferromagnetic super exchange modulated by proximity to the orbitally ordered to disordered transition.

The presence of metal-insulator (M-I) transitions in magnetic materials, particularly in colossal magnetoresistive (CMR) perovskites manganites [1-5], have been the focus of intensive study. The general formula for the latter is usually given as A$_{1-x}$B$_x$MnO$_3$, x being the doping level, A = rare-earth, viz., La, Pr, etc., while B = a divalent alkaline earth ion, typically Ca or Sr. Historically, the explanation of CMR in manganites was based on the concept of spin-dependent double exchange (DE) [6] in which a ferromagnetic (FM) interaction between the localized t$_{2g}$ spins resulted from the hopping of itinerant e$_g$ spins between adjacent Mn atoms subject to strong intra site Hund's rule coupling. A specific prediction of the conventional DE picture was that the onset of metallicity reflected the establishment of an infinite (percolating) pathway of DE metallic bonds, with these same bonds establishing an infinite FM "backbone", so that the emergence of metallicity and ferromagnetism were coincident. Current studies are at variance with such a prediction; in La$_{1-x}$Ca$_x$MnO$_3$ for example [4,7-11], for doping levels 0.18 ≤ x$_c$ ≤ 0.22 (i.e. carrier concentration n = 1-x), a FM ground state persists but is accompanied alternatively by conduct-ing (x ~ 0.22) or insulating (x ~ 0.18) characteristics. Recent experiments indicate that the emergence of an insulating, as opposed to a metallic, ground state reflects local structural changes (which control carrier (e$_g$) (de)localization [10]), characterized quanti-tatively by the absence/presence of the so-called Jahn-Teller (JT)-long-bond accompanied by an orbitally ordered (OO) (insulator) to orbitally disordered (OO*) (metal) transition [9, 11]. The origin of ferromagnetism in this composition range thus remains elusive, if not controversial. While quantitative correlations between JT distortions and both the transport and magnetic response across this compositional phase boundary have been established experimentally [9-11], differences in spin excitation processes, viz., the appearance of spin diffusive modes in metallic samples but their absence in the insulating regime, indicate that the associated magnetism *might* be different [8]. Theoretically, the emergence of a FM insulating phase is not a universal model prediction [12].

The present study attempts to address the failure of the DE model in its prediction for the coincident emergence of ferromagnetism and metallicity in the La$_{1-x}$Ca$_x$MnO$_3$ system via measurements of the critical exponents and amplitudes across the compositionally controlled M-I boundary. Data on two La$_{1-x}$Ca$_x$MnO$_3$ single crystals with x = 0.18 and 0.20, grown using the floating zone technique [13], are summarized below. Their high structural/magnetic qualities are confirmed by a mosaicity typically less than 1° and low coercive field. Measurements of the ac susceptibility, χ(H, T), (at 1 kHz with an ac driving field of 0.1 Oe rms) and magnetization, M(H, T), were carried out in a Quantum Design PPMS Model 6000 susceptometer/magnetometer, with all fields applied along the largest sample dimension to minimize demagnetization effects. Magnetoresesistivities, ρ(H, T), were acquired with a Model 7000 AC Transport Controller using a four-probe technique with an excitation current 10 µA at 499 Hz.

Figures 1a (x = 0.18) and 1b (x = 0.20) reproduce magnetic isotherms at selected temperatures; the inserts display the corresponding zero field ac susceptibilities, χ(0, T), measured (in 1 K steps) on warming (no hysteresis was detected between cooling and warming through the transition region; the structural phase transition near T$_B$ ≈ 50 K for 0.125 ≤ x ≤ 0.20 [7] not being a focus of the present study). The Hopkinson/principle maxima [14] evident in these inserts yield an estimate for the demagnetization factors (N$_D$), and the inflection points a preliminary determination of T$_C$ ≈ 171 K for x = 0.18 and T$_C$ ≈ 179 K for x = 0.20. Magnetization isotherms in the vicinity of these transition temperatures displays no "S" shaped features characteristic of a metamagnetic/discontinuous transition [15-17], the continuous nature of ferromagnetic-paramagnetic (FM-PM) transitions being confirmed by the positive slope of the corresponding Arrott plots [15]. These magnetization data are reproduced in the form (H$_i$/M)$^{1/1.387}$ vs M$^{1/0.365}$, Figs. 1c (x = 0.18) and 1d (x = 0.20), suggested by a modified Arrott-Noakes equation of state [16]; the resulting parallel lines confirm the applicability of near 3-D Heisenberg model exponents ($\gamma = 1.387$, $\beta = 0.365$, $\delta = 4.783$ [18]), and yield values of $T_C \approx 171 \pm 1$ K (x = 0.18) and $T_C \approx 179 \pm 1$ K (x = 0.20) from the critical isotherms. Along the latter, standard scaling theory predicts [14-17] $M(H_i, T = T_C) = D_0 H_i^{1/\delta}$, where $D_0$ is a critical amplitude, a prediction verified by the double-logarithmic plots inserted in these figures and which yield $\delta = 4.783 \pm 0.004$, $D_0 = 5.4 \pm 0.1$ emu/g Oe (x = 0.18), and $\delta = 4.785 \pm 0.003$, $D_0 = 5.2 \pm 0.1$ emu/g Oe (x = 0.20). These $\delta$ estimates are not only in close agreement with 3-D Heisenberg model predictions, but they also confirm – albeit indirectly – the absence of Griffiths phase-like features (GP) [19]. Separate tests of the power-law predictions for the inverse initial susceptibility, $1/\chi_i(T > T_C) = (\partial H/\partial M)_{H=0} = \chi_0|t|^\gamma$ and the spontaneous magnetization $M_S(H = 0, T < T_C) = M_S(0)|t|^\beta$ (where $|t| = |T/T_C-1|$) yields critical amplitudes $\chi_0 = (8.9 \pm 0.5) \times 10^3$ g-Oe/emu, $M_S(0) = 124 \pm 4$ emu/g (x = 0.18) and $\chi_0 = (1.11 \pm 0.06) \times 10^4$ g-Oe/emu, $M_S(0) = 132 \pm 3$ emu/g (x = 0.20). These power-law plots are not reproduced here, although a final assess-ment of scaling behavior (viz., $M(H_i,t) = |t|^\beta \cdot F_\pm[H_i/(|t|^{\beta\delta})]$ where $F_\pm(x)$ is the (unknown) scaling function above (+)/below(-) $T_C$ [14,17,19,20] is carried out in Fig. 1e using the listed exponent values, where the good data collapse evident confirm the applicability of 3-D Heisenberg exponents to both single crystals. Estimates of the low temperature spontaneous magnetization, $M_S$, obtained from the intercepts of plots similar to those shown in Figs. 1c and 1d, enable values for the acoustic spin wave stiffness, D, to be found using the well-established Bloch $T^{3/2}$ law [21]; fits to the latter indicate D ~ 65 mev Å$^2$ in both samples, consistent with the previous results on Ca doped samples showing no evidence of GP-like features [4,7,8,19]. Estimates of the saturation and spontaneous magnetizations, $M_{sat}$ and $M_S$, at 2 K have been used to verify the nominal composition of these samples by comparing the latter with the theoretically predicted spin-only moment of (4 - x) $\mu_B$/Mn, yielding $M_{sat}(T = 2$ K$) \approx M_S(T = 2$ K$) = 94.8 \pm 0.3$ emu/g $= 3.63 \pm 0.03$ $\mu_B$/Mn for x = 0.18 and $M_{sat}(T = 2$ K$) \approx M_S(T = 2$ K$) = 95.5 \pm 0.4$ emu/g $= 3.60 \pm 0.02$ $\mu_B$/Mn for x = 0.20. Both estimates suggest that here the $e_g$ electrons fully spin-polarized with no significant spin canting [8,22].

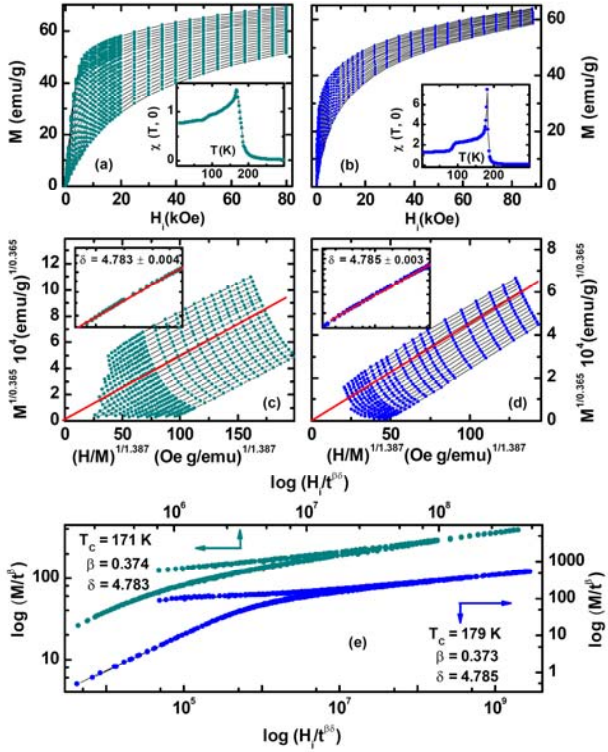

Figure 1. (Color online) Main bodies of (a) and (b) are a series of selected magnetic isotherms: (a) x = 0.18, collected in increasing field from 150 K to 190 K in 2 K steps; (b) x = 0.20, collected from 175 K to 195 K in 1 K steps. Insets to (a) and (b) are the zero field ac susceptibility $\chi(H = 0, T)$. (c) and (d) present a series of parallel lines, reproduced in the form $(H_i/M)^{1/1.387}$ versus $M^{1/0.365}$ using data in (a) and (b). The linear simulations to the isotherm passing through the origins yield $T_C$. Insets are plots of $\log(M(H_i, T = T_C))$ vs $\log(H_i)$, yielding $\delta$, using data in the range of 2 kOe < $H_i$ < 80 kOe. (e) Scaling plot of $\log(M/t^\beta)$ vs $\log(H_i/t^{\beta\delta})$, using the exponents and $T_C$ listed above. The upper branch corresponds to data below $T_C$ and the lower branch to data above $T_C$.

Ac susceptibility data are in complete agreement with the above conclusions. Figs. 2a (x = 0.18) and 2b (x = 0.20) show that in static biasing fields, $H_a$, a series of critical susceptibility maxima emerge, the temperature ($T_m$) of which increases while the amplitude ($\chi(H_a, T_m)$) decreases as $H_a$ is increased. The insets in these figures show the inverse zero-field ac susceptibilities ($1/\chi(0, T)$), in which the lack of a characteristic downturn just above $T_C$ provides convincing evidence regarding the absence of GP-like features [19] in these samples. The evolution of the critical maxima mentioned above, with field and temperature, in terms of scaling law predictions have been well documented previously [14,17], so that a simple summary of those predictions are given here. First, the dependence of the ac susceptibility peak amplitude on field, $\chi_m(H_i, t_m) \sim H_i^{(1/\delta-1)}$, is tested in the log-log plot of Fig. 2c, and yields $\delta = 4.783 \pm 0.003$ (x = 0.18) and $\delta = 4.779 \pm 0.004$ (x = 0.20) (estimates that are clearly independent of any choice for $T_C$). Second, the predicted power-law dependence of the (reduced) peak temperature on field, $t_m = (T_m - T_C)/T_C \sim H_i^{0.57} \sim H_i^{1/(\beta+\gamma)}$, is similarly reproduced in Fig. 2d; here an estimate for $T_C$ is clearly required, and is furnished by plotting the measured $T_m$ against $H_i^{0.57}$ and extrapolating to $H_i = 0$ (viz. $(\gamma+\beta)^{-1} = 0.57$ for Heisenberg model exponents [18]), yielding $T_C = 171 \pm 1$ K (x = 0.18) and $T_C = 179 \pm 1$ K (x = 0.20). The latter are then used in double-logarithmic plots of $t_m$ against $H_i^{1/(\beta+\gamma)}$ to yields self-consistently refined values [14,17] for $\gamma+\beta$ and $T_C$. Third, the dependence of peak amplitude on (reduced) temperature, $\chi_m(H_i, t_m) \sim t_m^{-\gamma}$, is similarly tested (Fig. 2e), yielding $\gamma = 1.379 \pm 0.005$ and hence $\beta = 0.374 \pm 0.002$ (x = 0.18), and $\gamma = 1.382 \pm 0.004$ and $\beta = 0.374 \pm 0.002$ (x = 0.20). A final test of the applicability of 3-D Heisenberg model is provided in Figs. 2g and 2h; scaling predicts that such $\chi(H, T)$, when normalized to its peak value ($\chi(h, T_m)$), should fall on a universal curve when plotted against the argument, $t/h^{1/(\gamma+\beta)}$ of the scaling function [14,17].

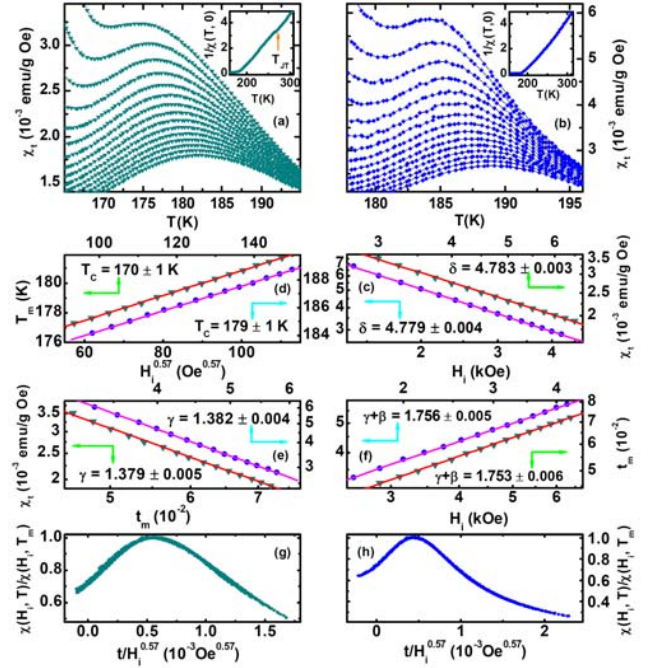

Figure 2: (Color online) Main bodies of (a) (x = 0.18) and (b) (x = 0.20) are $\chi(H, T)$ (corrected for background and demagnetizing effects) measured on warming following ZFC in various static fields, (a) from 5 kOe (top) to 9 kOe (bottom) in 250 Oe steps, (b) from 1.4 kOe (top) to 4.2 kOe (bottom) in 200 Oe steps. Insets are $1/\chi(H = 0, T)$. (c) Plot of $\log(\chi_m)$ vs $\log(H_i)$, yielding $\delta$. (d) Plot of $(T_m)$ against $(H_i^{0.57})$ yields estimate of $T_C$. (e) Plot of $\log(\chi_m)$ vs $\log(t_m)$, yielding $\gamma$. (f) Plot of $\log(t_m)$ vs $\log(H_i)$, yielding $\gamma + \beta$ and hence $\beta$. (g) (x = 0.18) and (f) (x = 0.20) are the susceptibility scaling using the data in (a) and (b).

That these samples straddle the compositional M-I boundary is confirmed by the (magneto)resistivity (MR) data, $\rho(H, T)$, reproduced as a function of temperature in various applied fields, $0 < H < 90$ kOe, in Figs. 3a (x = 0.18) and 3b (x = 0.20). These data support the conclusion that this compositional threshold lies in the range $0.18 \leq x_c \leq 0.22$ [4,5,8-11]. In particular, the x = 0.18 sample exhibits an insulating ground state ($\rho(H = 0, T = 10$ K$) = 116$ Ω-cm) above which there are three phase transitions; a structural transition at $T_B \approx 50$ K, a higher temperature FM-PM transition $\approx 172$ K, and a JT transition at T $\approx 270$K, also evident in the $1/\chi(0, T)$ vs T plot, corresponding to a pseudo-cubic to orthorhombic structural transition (coincident with a OO*-OO transition) [7-11]). In contrast, the x = 0.20 specimen exhibits a metallic ground state ($\rho(H = 0, T = 10$ K$) = 1.85 \times 10^{-4}$ Ω-cm, i.e. close to a factor of $10^6$ lower than at x=0.18) with the associated FM-PM (M-I) transition $\approx 185$ K, with a strong field dependent resistivity [5] characteristic of CMR systems. Despite marked differences in ground state properties, by examining their response in the PM regime it is possible to compare the high temperature coupling between the (partially) mobile $e_g$ electrons and localized $t_{2g}$ spins. This can be done by using the Ginzburg-Landau/mean-field prediction for the PM state, viz. $\Delta\rho(H, T)/\rho(0) = C(M(H, T)/M_{sat})^2$ [23] where C represents the effective coupling between $e_g$ conduction electrons and localized $t_{2g}$ spins, and depends on the ratio $J_H/W$ ($J_H$ being the Hund's coupling between itinerant and localized electrons and W the one-electron $e_g$ bandwidth). This prediction is confirmed in Fig. 3c for data immediately above $T_C$ (i.e. T $\geq 1.02T_C$), from which C = 1.77 for x = 0.18, considerably larger than the estimate of C = 0.47 at x = 0.20; such a trend is consistent with that observed in La$_{1-x}$Sr$_x$MnO$_3$ near the M-I transition and supports the conclusion that the value for C decreases as the hole concentration, n, is increased [5,23]. Adherence to this prediction at compositions displaying an insulating ground state indicates that even in the case of (partially) localized $e_g$ spins, viz. in the absence of a percolating backbone of DE-linked $e_g$ electron hopping pathways, magnetism and conduction remain strongly coupled in the PM regime.

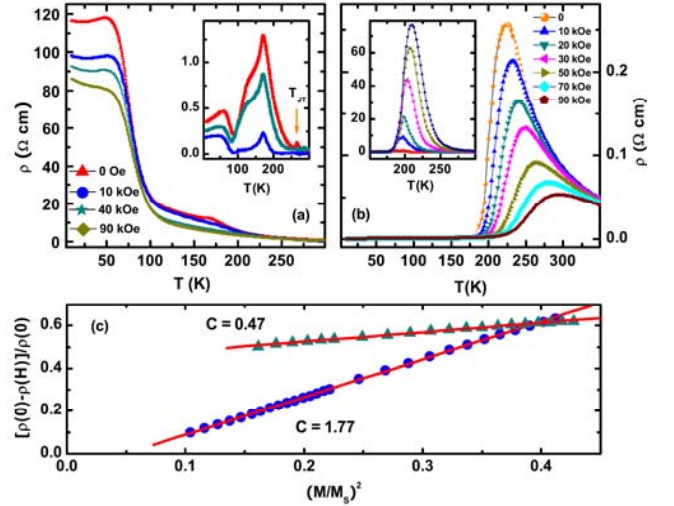

Figure 3. (Color online) (a) Resistivity for x = 0.18 measured on warming following ZFC; insert is the MR $[\rho(0) - \rho(H)]/\rho(0)$ exhibiting a peak around $T_C$. (b) as in (a) for x = 0.20 and presents a CMR behavior. (c) The scaling of MR with magnetization, $\Delta\rho/\rho(0) = C(M/M_{sat})^2$, using the data immediately above $T_C$.

The fundamental question raised by the present results, however, is what coupling mechanism underlies the establishment of a FM ground state across the compositional M-I transition, a transition characterized by the presence/absence of metallicity and hence a percolating hopping pathway of DE linked sites? Should the DE interaction play a principal role in spin-spin coupling on either side of this M-I transition, a clear corollary would be a significant increase in the associated critical amplitudes, $D_0$, $\chi_0$ and $M_S(0)$, on establishing metallicity. This does not happen. The implication then is that DE is not the principal mechanism underlying the establishment of ferromagnetism in this composition range.

We suggest that the relevant interaction is super exchange (SE), based on the evolution of the sign/magnitude of SE with composition deduced from neutron scattering data [24]. The parent compound, LaMnO$_3$, exhibits "in-plane" ferromagnetism, reflecting the fact that the corresponding in-plane SE coupling integral $J_{ab} > 0$, and out-of-plane antiferromagnetic (AFM) SE coupling $J_c < 0$, a magnetic structure induced/stabilized by OO [5,25] ( this structure can also be discussed in terms of the semi-empirical Kanamori-Goodenough-Anderson (KGA) rules [5,25], in which the sign of SE coupling reflects the Mn-O-Mn bond angles and bond lengths). These neutron data indicate that whereas the magnitude of the FM-SE in-plane coupling $J_{ab}(x)$ increased monotonically with increasing doping level, x, over the relevant range, the c-axis coupling $J_c(x)$ evolved in a more complicated manner, specifically, $J_c(x < 0.125) < 0$, $J_c(x = 0.125) = 0$, while $J_c(0.125 < x < 0.22) > 0$ − indicating the emergence of a FM-SE c-axis coupling for $x > 0.125$, and subsequently increasing roughly linearly with x (for $x > 0.125$) [24]. This demonstrates not only that both $J_{ab}(x)$ and $J_c(x)$ are positive/FM across the compositionally induced M-I transition in $La_{1-x}Ca_xMnO_3$, but also, should these SE interactions dominate the ordering process – as suggested above – then, (i) $T_C$ would increase smoothly with composition across the M-I boundary, as observed (a much sharper increase would be expected in the DE dominated counterpart accompany the emergence of an infinite percolating metallic pathway), (ii) the critical amplitudes should also exhibit little change across the same boundary, again as observed (again contrasting with DE model expectations), and (iii), the universality class for the transition would necessarily remain the same. While Monte Carlo simulations conclude that the DE model - without anisotropy – lies in the same universality class as the Heisenberg model [26], no such predictions currently exist for SE; the present measurements reveal that it, too, lies in the same class. In support of the above, several "model" phase diagrams expectedly designate the zero-temperature M-I critical composition, $x_c$, as that at which the (insulating) OO regime is replaced by the (metallic) OO* disordered phase [11]. In the vicinity of the OO–OO* boundary considerable inhomogeneity – likely electronic phase separation [11,27] – is anticipated, with the present data showing that such inhomogeneity, albeit measured indirectly by the coercive field (at 10 K, $H_C = 35$ Oe for $x = 0.18$ while $H_C = 8$ Oe at $x = 0.20$), is, perhaps not unexpectedly, weaker following the establishment of an infinite, percolating metallic backbone; the latter appears qualitatively consistent with the observation of "confined" spin waves in the composition region below $x_c$ [24,27].

In summary, measurements of the critical exponents and amplitudes in two single crystal $La_{1-x}Ca_xMnO_3$ spanning the compositionally driven M-I transition are presented. The similarities in these critical parameters demonstrate that the establishment of an infinite, percolating metallic backbone of DE linked sites has no measurable influence on the accompanying magnetic transition, from which it is inferred that DE is not the principal mechanism underlying the establishment of a FM ground state in this compositional regime. Arguments are presented that FM-SE evolving in the vicinity of the (insulating) OO to (metallic) OO* boundary underlies this behaviour, and subsequently that such FM-SE interactions lie in the same universality class as the 3-D Heisenberg model. Indirect support for this latter conclusion is provided by the use of the same model to fit the measured dispersion relation for magnetic excitations for $x < x_c$ [24]. The similarity in magnetic behaviour reported in the above is not confined to the critical region, but extends well throughout the ordered phase, as manifested by comparable values of the acoustic spinwave stiffness, D, in samples with compositions on either side of the M-I transition.

We thank the helpful discussions with J. Neumeier, and A. Millis, J. S. Zhou. Support for this work by the Natural Sciences and Engineering Research Council (NSERC) of Canada and the University of Manitoba (to WJ) is gratefully acknowledged.

jiang@physics.umanitoba.ca